\documentclass[twocolumn]{aastex631}

\submitjournal{ApjL}
\usepackage{units}
\usepackage{mathtools}

\begin{document}

\title{On The Theory of Ring Afterglows}

\author[0000-0003-3356-880X]{Marcus DuPont}
\author[0000-0002-0106-9013]{Andrew MacFadyen}
\affiliation{Center for Cosmology and Particle Physics, New York University\\
New York, NY 10003, USA}

\author[0000-0002-1084-3656]{Re'em Sari}
\affiliation{Racah Institute, Hebrew University of Jerusalem \\
Jerusalem, 91904, Israel}



\begin{abstract}
Synchrotron and inverse Compton emission successfully explain the observed spectra of gamma-ray burst (GRB) afterglows. It is thought that most GRBs are products of extremely relativistic outflows and the afterglow marks the interaction of that ejecta with the surrounding matter. Faster decay of afterglow light curves at late times is indicative  of non-spherical geometries, and are usually interpreted as evidence for jet geometry. Recent numerical simulations have shown that ring-like geometries are also permissible for relativistic outflows. We therefore extend the standard theory of afterglow evolution to ring geometries. An analytic prescription for the light curves and spectra produced by relativistic toroidal blast waves is presented. We compare these to their spherical and jet-like counterparts, and show that ring afterglows decay faster than spherical outflows but not as fast as jets. 
\end{abstract}



\section{Introduction} \label{sec:intro}
The physics accounting for the variability and wide range in observed luminosities of gamma-ray bursts (GRBs), and the nature of their central engine  are topics of deep debate. However, it is widely accepted that the dominant processes responsible for the X-ray, optical, and radio afterglow radiation are the synchrotron and inverse Compton mechanisms operating behind the blast wave that the GRB launches into the surrounding medium.
Such radiative processes are expected to be applicable to
just about any sufficiently relativistic outflow. This paved the way for the success of using the \cite{Blandford+Mckee+1976}  (BM) solution for modelling GRB afterglows and for distinguishing between isotropic and jet-like asymmetric outflows modelled as BM solutions truncated to within polar angle $\theta_0$ \citep[see][and references therein]{Piran+2004}. Thus far, only afterglows for spherical and jet-like outflows have been considered and it is generally believed that most GRBs are caused by jetted relativistic outflows. Currently, the key indicators cited as evidence for GRB jets are: (a) the existence of an achromatic break in the afterglow light curve either due to lateral jet spreading \citep{Rhoads+1999, Sari+Piran+Halpern+1999} or an off-axis viewing of universal structured jets \citep[e.g.,][]{Zhang+2002, Rossi+2002}; (b) observed net polarizations that arise from asymmetric, relativistically beamed outflows \citep[e.g., ][]{Gruzinov+Waxman+1999, Sari+1999, Yonetoku+2011, Mandarakas+2023};
(c) extremely large energetics which require the outflow to be sufficiently collimated since the average \emph{isotropic} energy of $\unit[10^{55}]{erg}$ released by GRBs is much larger than what is physically allowed by a spherical explosion of a massive star \citep{Taylor+2004,Kumar+zhang+2015}; (d) and measurements of proper motion of the flux centroid \citep{Czerny+1997,Taylor+2004,Mooley+2018}.

Insofar as shown by the the previous conditions and observations, many GRBs are only constrained to be \emph{asymmetric} outflows, but  we argue they are not necessarily jet-like. This stance is valid since the current GRB afterglow catalogue is quite varied and many of them show breaks which do not fit the jet theory.  Recently, it has been shown that relativistic outflows can have ring-like geometries, e.g. from the ``ellipsar'' mechanism \citep{DuPont+2022}. Motivated by the result of \citet{DuPont+2022}, we 
consider in this Letter the dynamics and observational signatures
of expanding relativistic rings,
though we remain agnostic about the source and energies of said rings. 
Our work on ring afterglows is motivated
by the many time-domain surveys in progress or being planned \citep{Barthelmy+2005, Shappee+2014,Chambers+2016,Kochanek+2017,Ivezi+2019,Bellm+2019}, which observe a wide array of astrophysical transients --- outside of just GRBs --- that expanding relativistic rings might help explain. These transients might include X-ray flashes (XRFs), Super Luminous Supernovae (SLSNe), trans-relativistic supernovae, and Fast Blue Optical Transients (FBOTs). Therefore, we are motivated to ask how the afterglow of expanding relativistic rings differs from their spherical and jet-like counterparts.  

In this Letter, we calculate the light curves and spectra due to expanding relativistic rings. We invoke the same recipes
described in \cite{Sari+Piran+Narayan+1998} and \cite{Sari+Piran+Halpern+1999}, which have been successful at modeling many observed GRB afterglows. We derive temporal scalings for the relevant frequencies and spectral flux and comment on their differences from the spherical and jet-like afterglows. 

This Letter is organized as follows: Section \ref{sec:formalism} describes the mathematical formalism for the dynamics and synchrotron radiation from the relativistic ring, Section \ref{sec:lightcurves} describes the resultant light curves of the ring-like outflows, and Section \ref{sec:discussion} discusses the relevance of our work. 

\section{Formalism} \label{sec:formalism}
\subsection{Blast wave evolution}
\begin{figure*}
    \plottwo{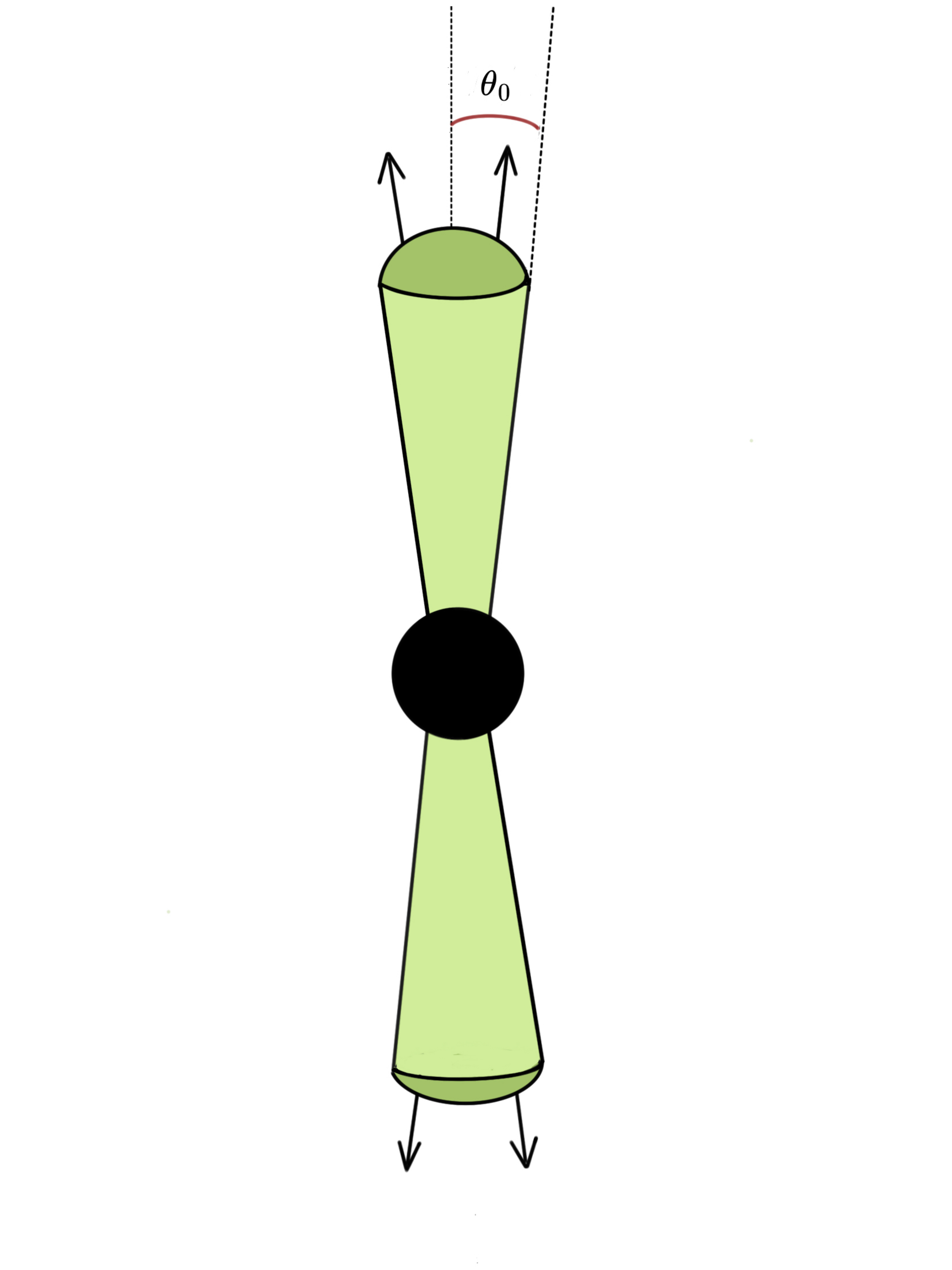}{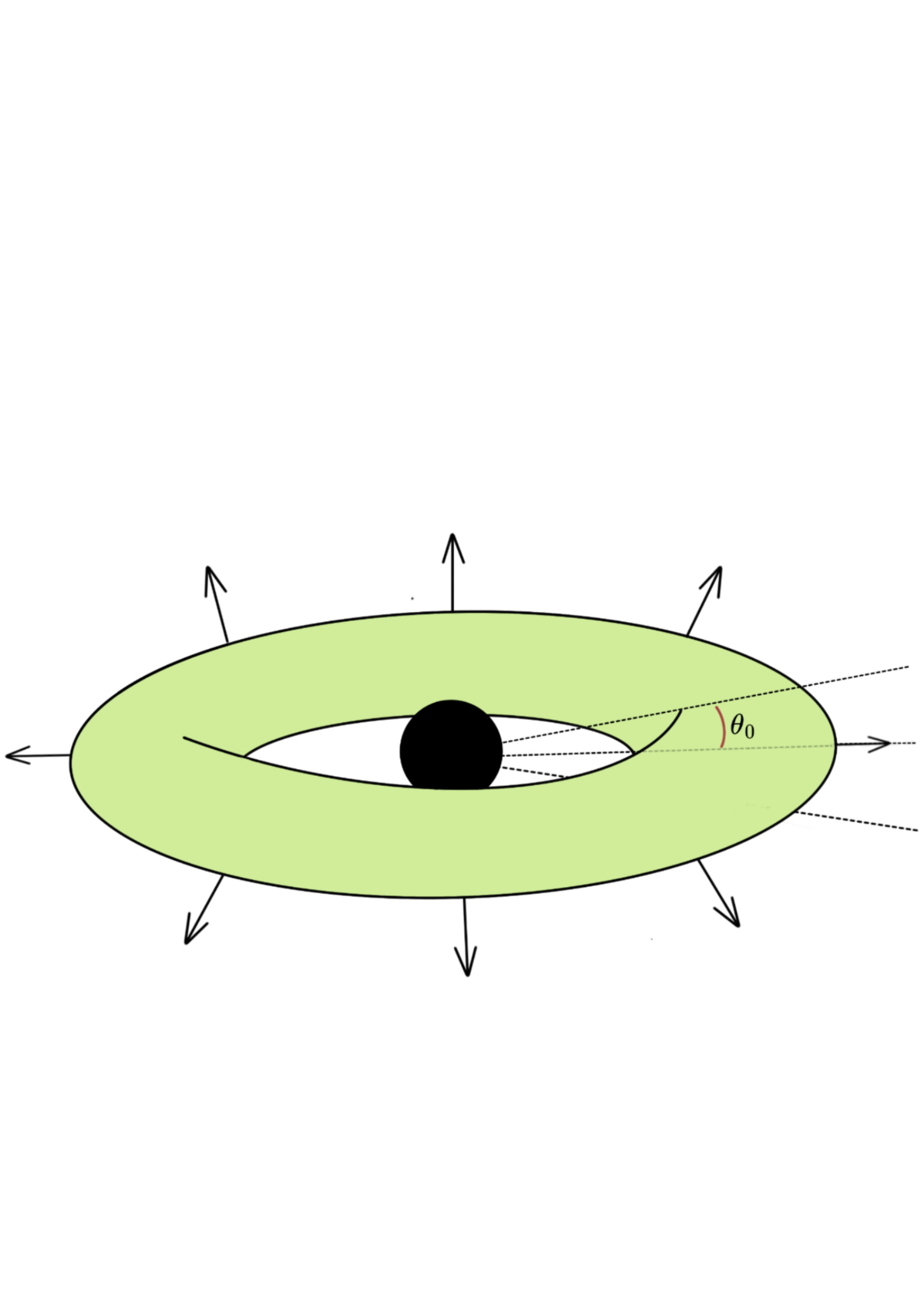}
    \caption{Cartoon illustrations of the two types of asymmetric geometries considered in this Letter. The left shows the conical, jet-like outflow along the poles of the source while the right shows the ring-like outflow in the equatorial plane of the source. The half-opening angle, $\theta_0$, is depicted for both geometries as well.
    }
    \label{fig:cartoon}
\end{figure*}

In the early phase of evolution before the expanding blast wave begins to decelerate, if it is expanding in a medium with density obeying $\rho = A r^{-k}$, it has kinetic energy
\begin{equation}\label{eq:energy}
    E \approx \Gamma^2 M = \frac{A}{3 -k} \Gamma^2 r^{3 - k}\Omega,
\end{equation}
where $M$ is the swept up mass, $\Gamma = (1 - \beta^2)^{-1/2}$ is the Lorentz factor of the bulk flow, $\beta$ is velocity in units of $c$, $A$ is the mass-loading parameter, and $\Omega$ is the solid angle of the blast wave which obeys
\begin{equation}\label{eq:solid_angle}
    \Omega = 
    \begin{cases}
        4\pi\sin(\theta_0) &\ \text{ring}, \\
        8\pi\sin^2(\theta_0 / 2) &\ \text{jets}, \\
        4\pi & \ \text{sphere,}
    \end{cases}
\end{equation}
where $\theta_0$ is the half-opening angle of the blast wave(s) such that $\Omega \rightarrow 4\pi$ as $\theta_0 \rightarrow \pi / 2$. For small opening angles, $\Omega_{\rm ring} \approx 4\pi\theta_0$, which is a factor $2 /\theta_0$ larger than its double-sided jet-like counterpart, making relativistic rings more likely to be observed, as compared to a jet with the same opening angle. An illustration of the asymmetric geometries considered is shown in Figure \ref{fig:cartoon}. 
As evident from Figure \ref{fig:cartoon} and from Equation \ref{eq:solid_angle}, the solid angle for a ring complements the solid angle of a jet to $4\pi$ if $\theta_{\rm ring}=\pi/2-\theta_{\rm jet}$.

Using conservation of energy, as the relativistic ring slows down such that $\Gamma \sim \theta_0^{-1}$, one finds $\Gamma \propto r^{-(3-k)}$. This happens after an observer time of:
\begin{equation}\label{eq: tbreak}
    t_{\rm b} \approx 
    [\zeta E_{\rm iso} / 4\pi A]^{1/\zeta}(\theta_0 + \theta_{\rm obs})^{2(1+\zeta)/\zeta},
\end{equation}
where $E_{\rm iso}$ is the isotropic-equivalent energy and $\zeta \equiv 3-k$. Before this break time, the afterglow from rings and from jets are identical due to a lack of causal connectivity. The crux of this Letter is that after this break time, light curves from jets and from rings diverge and their distinguishing features are discernible in the current GRB catalogue. We will explore the previous point in later sections. 

As the blast wave evolves, an observer sees photons at a time 
\begin{equation}\label{eq:observer_time}
    t = t^\prime(1 - \vec{\beta} \cdot \hat{n}) = t^\prime (1 - \beta \mu),
\end{equation}
where $t^\prime$ is the time in the emitter frame, $\hat{n}$ is a unit vector pointing from the observer to the emitting patch, and $\mu \equiv \cos \theta$. Hereafter, all primed quantities signify values in the emitter frame. Assuming $\Gamma \gg 1$ and the observer is nearly perfectly oriented with the emitting patch (i.e., $\mu \approx 1 - \theta^2/2$), we have 
\begin{equation}\label{eq:observer_time_radius}
    t \approx \frac{t^\prime}{2\Gamma^2}[1 + (\Gamma \theta)^2] \approx \frac{r}{2\Gamma^2}[1 + (\Gamma \theta)^2],
\end{equation}
where we have used $t^\prime \approx r$ for sufficiently relativistic flows which lie on the light sphere. Since the radiation is beamed into a typical angle $1  / \Gamma$, the quantity $\Gamma \theta$ is of order unity, simplifying the observer time to $t \approx r / \Gamma^2$. From this, we arrive at the Lorentz factor as a function of observer time for the ring, $\Gamma \propto t^{-\zeta / (1 + 2\zeta)}$. Furthermore, the relativistic ring's radial evolution obeys $r \propto t^{1/(1 + 2\zeta)}$ after spreading begins.

%
\subsection{Synchrotron Spectrum}
In the observer frame, the characteristic peak frequency of the electrons is
\begin{equation}\label{eq:num}
    \nu_{m} = \Gamma \gamma_e^2 \frac{3eB'}{16 m_e} \propto \Gamma^4 \propto t^{-4\zeta / (1 + 2\zeta)},
\end{equation}
where $\gamma_e$ is the electron Lorentz factor, $e$ is elementary charge, $B^\prime$ is the magnetic field in the fluid frame, and $m_e$ is the electron mass. Note that we have used the fact that the magnetic field in the down stream transforms from the usual jump condition $B^\prime = \Gamma \sqrt{32\pi \rho \epsilon_B}$, where $\epsilon_B$ is fraction of total energy density due to magnetic fields, and the minimum Lorentz factor of the electrons obeys $\gamma_e \propto \Gamma$. 
In a time $t^\prime$, the electrons cool at a rate $\frac{\Gamma \gamma_e m_e}{t^\prime} = \langle P(\gamma_e) \rangle$, where $\langle P(\gamma_e) \rangle$ is the average power per electron per pitch angle \citep{Rybciki+Lightman+1979}:
\begin{equation}\label{eq: avg_power}
    \langle P(\gamma_e) \rangle =  \frac{4}{3}\sigma_T u^2 \gamma_e^2 U_b.
\end{equation}
In the above equation, $\sigma_T$ is the Thompson cross section, $[u^\mu] = \Gamma(1, \vec{\beta})$ is the four-velocity in units where $c = 1$, and $U_b = B^2 / 8\pi$ is the magnetic energy density. By inverting Equation \ref{eq: avg_power}, we solve for the cooling Lorentz factor,
\begin{equation}\label{eq:cooling_gamma}
    \gamma_c = \frac{6\pi m_e}{\Gamma t^\prime \sigma_T B^2} = \frac{6\pi m_e}{\Gamma^3 t \sigma_T B^2}.
\end{equation}
It then immediately follows that the cooling frequency obeys
\begin{equation}
    \nu_c = \Gamma \gamma_c^2\frac{3eB^\prime}{16 m_e} \propto \Gamma^{-4}t^{-2} \propto t^{-2 / (1 + 2\zeta)}.
\end{equation}
The spectral flux from a radiating blast wave is given by 
\begin{equation}\label{eq:general_fnu}
    F_\nu = \frac{1 + z}{4\pi d_L^2} \int_{V} \delta^2 j_\nu^\prime d^3\vec{x},
\end{equation}
where $z$ is redshift, $d_L$ is luminosity distance, $\delta = 1 / \Gamma(1 - \vec{\beta} \cdot \hat{n})$ is the Doppler beaming factor with respect to the observer, and $j_\nu^\prime$ is the frequency-dependent emissivity. At peak emission, the emissivity is independent of $\Gamma$ and a highly relativistic flow along the line of sight to the observer gives $\delta = 2\Gamma$, so the peak spectral flux has the scaling
\begin{equation}\label{eq:flux_scaling}
    F_{\nu, \rm max} \propto r^3\Gamma^2 \propto  t^{(3 -2\zeta)/(1 + 2\zeta)}.
\end{equation}
 For completeness, we do not assume that all synchrotron photons escape the plasma on their way to the observer, meaning some are self absorbed. Moreover, the self-absorption frequency is a difficult calculation, but by extrapolating from the \cite{Granot+et+al+1999} solution we can arrive at the simple scaling,
\begin{equation}\label{eq:nu_ssa}
    \nu_a \propto E^{1/5} \propto \Gamma^{2/5}r^{\zeta/5} \propto  t^{-\zeta/5(1 + 2\zeta)}.
\end{equation}
From this, we now have the necessary ingredients to compute light curves produced by relativistic rings.

\section{Light Curves of Relativistic Rings}\label{sec:lightcurves}
With the necessary constraints derived in the previous section, we now turn to explicit light curve calculations. Hereafter, we compute light curves for a constant density medium (i.e., $\zeta = 3$) to easily compare with the spherical and jet-like geometries derived in \cite{Sari+Piran+Halpern+1999}. We start with the flux at low enough frequencies, such that some photons are self absorbed. Assuming that the time-averaged source emits at the characteristic $\nu_m$, if $\nu_a \ll \nu_m$, then because most of the electrons are emitting at typical synchrotron frequencies much larger than $\nu_a$, the spectral flux is proportional to $\nu^2$ as opposed to $\nu^{5/2}$~\citep{Katz+1994}. Thus, we have 
\begin{widetext}
\begin{align}
    F_{\nu < \nu_a} \propto \left(\frac{\nu}{\nu_a} \right)^2 \left(\frac{\nu_a}{\nu_m} \right)^{1/3} F_{\nu, \rm max} \propto r^2 &\propto
    \begin{cases}
        t^{2/7} &\ \text{ring},\\
        \text{constant} &\ \text{jet},\\ 
        t^{1/2} &\ \text{spherical},
    \end{cases}\label{eq: flux_ssa}\\
%
%
    F_{\nu_a < \nu < \nu_m} \propto \left(\frac{\nu}{\nu_m} \right)^{1/3}F_{\nu, \rm max}
    \propto r^3 \Gamma^{2/3} &\propto
    \begin{cases}
        t^{1/7} &\ \text{ring},\\
        t^{-1/3} &\ \text{jet},\\ 
        t^{1/2} &\ \text{spherical},
    \end{cases}\label{eq: flux_intermed}
\end{align}
\end{widetext}
for the flux below the self-absorption frequency and the intermediate flux between the self-absorption and characteristic frequency, respectively. This indicates that slopes would rise as long as the evolution were spherical or ring-like, but the slopes are different enough to perfectly distinguish between the two geometries. Moreover, there is a stark contrast from the latter geometries when compared with the $t^{-1/3}$ decay of the jet once it begins spreading. At high frequencies, the light curves follow
\begin{figure*}[t]
    \plottwo{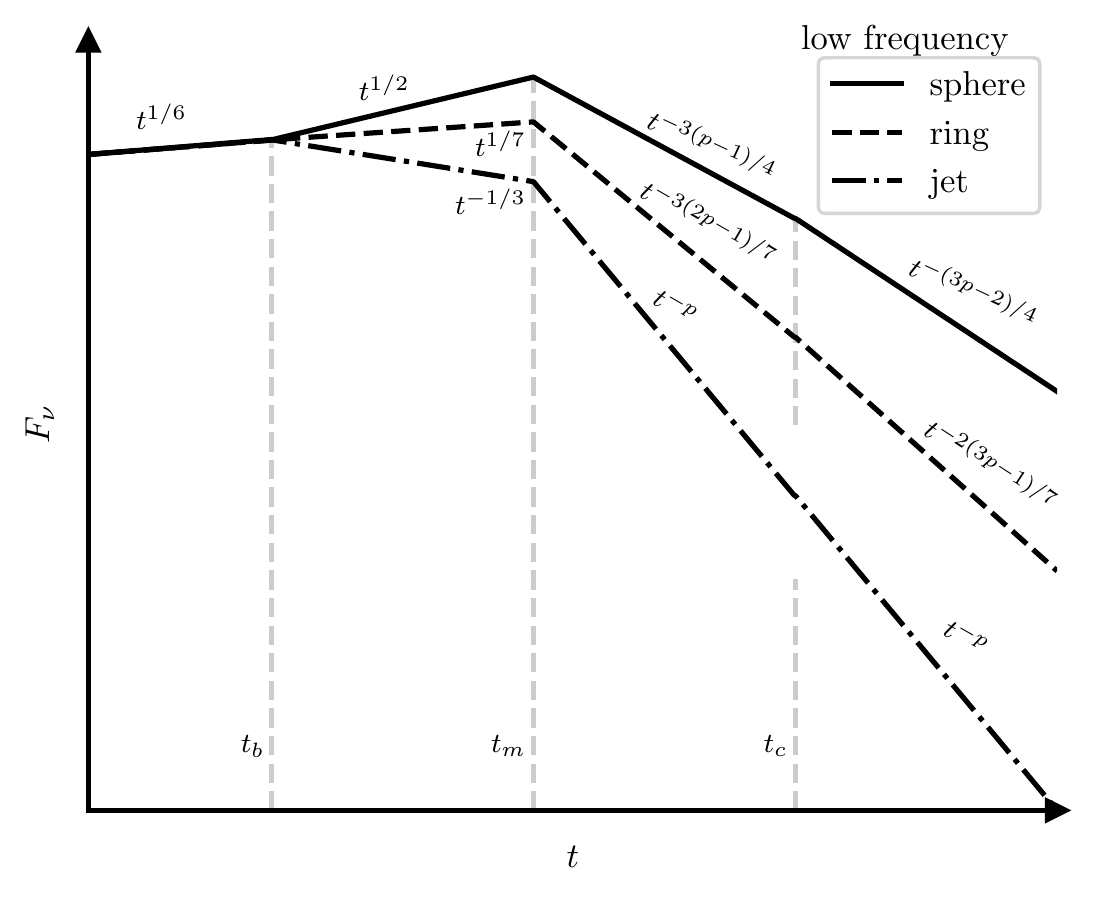}{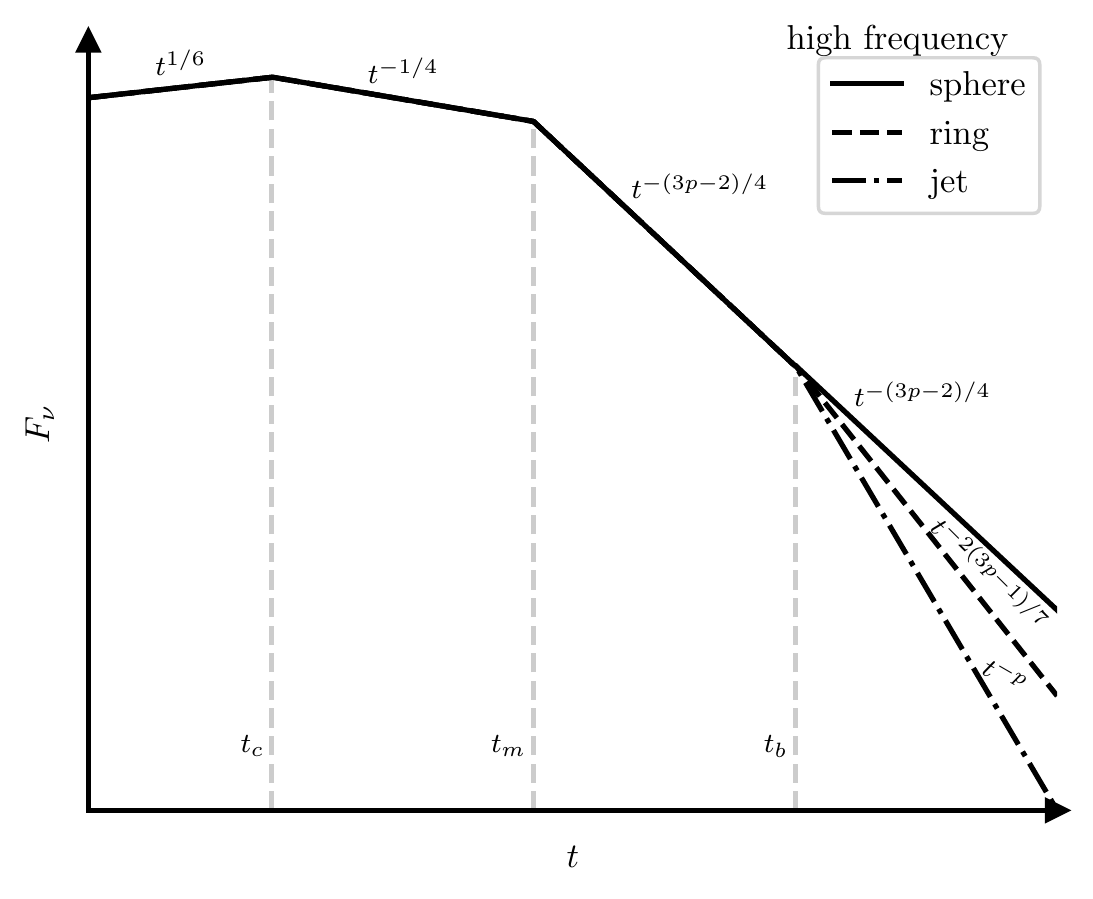}
    \caption{Pictorial light curves for the spherical, ring-like, and jet-like blast waves, respectively. The left and right panels show the typical light curve behavior at low ($\sim$ radio) and high ($\sim$ optical and X-ray) observed frequencies, respectively. The slopes are segmented in time between the break time, $t_b$, and the times when the break frequencies $\nu_m$ and $\nu_c$ cross the observed frequency $\nu$. The vertical dashed line for $t_c$ is broken at the jet curve in the left panel since $\nu_c$ is constant for that geometry and it therefore has no corresponding $t_c$.  In both frequency bands, we show the divergence in flux decay rate once the break time is reached with the low frequency band showing the clearest separation between the various phases of evolution.}
    \label{fig:light_curves}
\end{figure*}
\begin{widetext}
    \begin{align}
    F_{\nu_m < \nu_c < \nu} \propto \Gamma^2 r^3 \left(\frac{\nu_c}{\nu_m} \right)^{-(p-1)/2}\left(\frac{\nu}{\nu_c}\right)^{-p/2} \propto 
    \begin{cases}
        t^{-2(3p-1)/7} &\ \text{ring},\\
        t^{-p} &\ \text{jet},\\ 
        t^{-(3p - 2)/4} &\ \text{spherical},
    \end{cases}\label{eq: flux_cooling}\\
%
%
     F_{\nu_m < \nu <\nu_c} \propto \Gamma^2 r^3 \left(\frac{\nu}{\nu_m} \right)^{-(p-1)/2} \propto 
    \begin{cases}
        t^{-3(2p-1)/7} &\ \text{ring},\\
        t^{-p} &\ \text{jet},\\ 
        t^{-3(p - 1)/4} &\ \text{spherical},
    \end{cases}\label{eq: flux_non_cooling}
\end{align}
\end{widetext}
for cooling electrons and for non-cooling electrons, respectively. In Equations \ref{eq: flux_cooling} \& \ref{eq: flux_non_cooling}, $p$ is the electron distribution power-law index.
Here we witness that ring afterglows possess two distinct cooling breaks analogous to whats been calculated for spherical outflows. Furthermore, our calculation evidences a very clear distinction between afterglows produced by relativistic rings and jets. A graphical depiction of this distinction is shown in Figure \ref{fig:light_curves}. We've shown  \emph{very} distinct features such as differences in cooling breaks, and, more importantly, ring afterglows have shallower decay slopes than jets throughout the entirety of their evolution. The consequences of these revelations are discussed in the next section.

\section{Discussion}\label{sec:discussion}
We have demonstrated that temporal evolution of ring afterglows is clearly 
distinct from their spherical and jet-like counterparts.
While it is likely that classical GRBs are products of very energetic asymmetric flows, the geometry of said outflow is not well constrained. The jet model has been instrumental in its explanations of steep decays as resulting from highly collimated outflows. Yet, there exist observations which cannot be fit using the jet framework.  Some light curves --- such as those produced by GRB 030329 \citep{Stanek+03} or the more recent GRB 221009A \citep{Williams+23} --- have very shallow breaks, which are hard to reconcile using top-hat jet models. In particular, GRB 221009A was reported by \citet{Williams+23} to favor a broken power-law model in the X-ray with flux decay slopes steepening from $t^{-1.498 \pm 0.004}$ to $t^{-1.672 \pm 0.008}$ with a jet break time of $t_{b, \rm X-ray} \sim \unit[8 \times 10^4]{s}$. The timing of such steepening might be due to a jet with half-opening angle of $3.5^\circ$ \citep{Davanzo+2022}. However, the light curve does not steepen beyond the decay index $\alpha \approx 1.7$ --- where $F_\nu \propto t^{-\alpha}$ ---  after the break, and observers cannot match this shallow X-ray decay index with what is predicted using a simple on-axis top-hat jet. For typical values $p \cong 2.4$, the top-hat jets predict $\alpha > 2$, but rings predict $1.63 < \alpha < 1.77$, well within the required range for GRB 221009A. Therefore, one can interpret this GRB as stemming from either a more structured jet configuration or an expanding relativistic ring. 

The notion of some astrophysical transients being sourced from expanding relativistic rings, rather than jets, have the following implications: (a) the probability of viewing ring afterglows is larger than that of jets by a factor $2 / \theta_0$.
A blast wave with half-opening angle 0.1 radians, if oriented as a jet, would cover 0.5\% of the sky while an expanding ring covers 10\%, larger by a factor of 20.
As a result, ring geometries, as compared to jet geometries, bring down the required source rates significantly; (b) as demonstrated by \cite{DuPont+2022} relativistic rings can be born purely from  geometrical and hydrodynamic effects as opposed to the more complex central engines required for producing classical jets; (c) the late-time evolution of the relativistic ring is much more stable than the jet since the spreading of the relativistic ring is effectively one dimensional and is therefore a candidate for light curves with shallower breaks (d) around the time of the ring break, when the emitting patch is no longer locally spherical, the specific intensity (surface brightness) is no longer symmetric about the line of sight to the observer for a general viewing angle \citep{Granot+Piran+Sari+1999,Granot+Sari+Piran+1999, Granot+2007,vanEerten+2010}. This fact can act as useful probe for the underlying dynamics of rings which can be further detailed by direct hydrodynamic simulations in the near future. Detailed analysis of scintillation patterns may be sensitive to the surface brightness distribution \citep{Goodman+1997} and may help distinguish jets from rings.

This Letter has considered synchrotron emission which is more readily observed in radio and optical frequencies. At higher frequencies, inverse Compton emission may dominate. Adding the inverse Compton component could be done in a similar way to \citet{Sari+Esin+2001}, or its extension by \cite{Nakar+2009} if Klein Nishina corrections are important.

In summary, we've considered the dynamics and observational signatures of expanding relativistic rings which preserve the notion of beaming, can account for shallow breaks as observed in many GRB light curves, and do not require a complex central engine to achieve their geometric configuration. Our investigation is inspired by the work of \cite{DuPont+2022}, where rings arise naturally, albeit with lower energy than needed to explain cosmological afterglows for the conditions considered in that work. 
Moreover, while the main focus of this work has been GRBs, we emphasize that the importance of our calculations are the unique features presented by the ring geometry, while the energetics can be scaled appropriately and applied to a broader scope of astrophysical transients. Therefore, we suggest that ring-like outflows should be considered when interpreting observations of non-spherical explosions.  

\begin{acknowledgments}
M.D. thanks a James Arthur Fellowship from NYU's Center for Cosmology and Particle Physics and a KITP Graduate Fellowship, A.M. thanks NASA ATP grant 80NSSC22K0822, R.S. thanks ISF, MOST and BSF for support.
\end{acknowledgments}
%
%



\bibliography{refs}{}

\begin{thebibliography}{}
\expandafter\ifx\csname natexlab\endcsname\relax\def\natexlab#1{#1}\fi
\providecommand{\url}[1]{\href{#1}{#1}}
\providecommand{\dodoi}[1]{doi:~\href{http://doi.org/#1}{\nolinkurl{#1}}}
\providecommand{\doeprint}[1]{\href{http://ascl.net/#1}{\nolinkurl{http://ascl.net/#1}}}
\providecommand{\doarXiv}[1]{\href{https://arxiv.org/abs/#1}{\nolinkurl{https://arxiv.org/abs/#1}}}

\bibitem[{{Barthelmy} {et~al.}(2005){Barthelmy}, {Barbier}, {Cummings},
  {Fenimore}, {Gehrels}, {Hullinger}, {Krimm}, {Markwardt}, {Palmer},
  {Parsons}, {Sato}, {Suzuki}, {Takahashi}, {Tashiro}, \&
  {Tueller}}]{Barthelmy+2005}
{Barthelmy}, S.~D., {Barbier}, L.~M., {Cummings}, J.~R., {et~al.} 2005, \ssr,
  120, 143, \dodoi{10.1007/s11214-005-5096-3}

\bibitem[{{Bellm} {et~al.}(2019){Bellm}, {Kulkarni}, {Graham}, {Dekany},
  {Smith}, {Riddle}, {Masci}, {Helou}, {Prince}, {Adams}, {Barbarino},
  {Barlow}, {Bauer}, {Beck}, {Belicki}, {Biswas}, {Blagorodnova}, {Bodewits},
  {Bolin}, {Brinnel}, {Brooke}, {Bue}, {Bulla}, {Burruss}, {Cenko}, {Chang},
  {Connolly}, {Coughlin}, {Cromer}, {Cunningham}, {De}, {Delacroix}, {Desai},
  {Duev}, {Eadie}, {Farnham}, {Feeney}, {Feindt}, {Flynn}, {Franckowiak},
  {Frederick}, {Fremling}, {Gal-Yam}, {Gezari}, {Giomi}, {Goldstein},
  {Golkhou}, {Goobar}, {Groom}, {Hacopians}, {Hale}, {Henning}, {Ho}, {Hover},
  {Howell}, {Hung}, {Huppenkothen}, {Imel}, {Ip}, {Ivezi{\'c}}, {Jackson},
  {Jones}, {Juric}, {Kasliwal}, {Kaspi}, {Kaye}, {Kelley}, {Kowalski},
  {Kramer}, {Kupfer}, {Landry}, {Laher}, {Lee}, {Lin}, {Lin}, {Lunnan},
  {Giomi}, {Mahabal}, {Mao}, {Miller}, {Monkewitz}, {Murphy}, {Ngeow},
  {Nordin}, {Nugent}, {Ofek}, {Patterson}, {Penprase}, {Porter}, {Rauch},
  {Rebbapragada}, {Reiley}, {Rigault}, {Rodriguez}, {van Roestel}, {Rusholme},
  {van Santen}, {Schulze}, {Shupe}, {Singer}, {Soumagnac}, {Stein}, {Surace},
  {Sollerman}, {Szkody}, {Taddia}, {Terek}, {Van Sistine}, {van Velzen},
  {Vestrand}, {Walters}, {Ward}, {Ye}, {Yu}, {Yan}, \& {Zolkower}}]{Bellm+2019}
{Bellm}, E.~C., {Kulkarni}, S.~R., {Graham}, M.~J., {et~al.} 2019, \pasp, 131,
  018002, \dodoi{10.1088/1538-3873/aaecbe}

\bibitem[{{Blandford} \& {McKee}(1976)}]{Blandford+Mckee+1976}
{Blandford}, R.~D., \& {McKee}, C.~F. 1976, Physics of Fluids, 19, 1130,
  \dodoi{10.1063/1.861619}

\bibitem[{{Chambers} {et~al.}(2016){Chambers}, {Magnier}, {Metcalfe},
  {Flewelling}, {Huber}, {Waters}, {Denneau}, {Draper}, {Farrow}, {Finkbeiner},
  {Holmberg}, {Koppenhoefer}, {Price}, {Rest}, {Saglia}, {Schlafly}, {Smartt},
  {Sweeney}, {Wainscoat}, {Burgett}, {Chastel}, {Grav}, {Heasley}, {Hodapp},
  {Jedicke}, {Kaiser}, {Kudritzki}, {Luppino}, {Lupton}, {Monet}, {Morgan},
  {Onaka}, {Shiao}, {Stubbs}, {Tonry}, {White}, {Ba{\~n}ados}, {Bell},
  {Bender}, {Bernard}, {Boegner}, {Boffi}, {Botticella}, {Calamida},
  {Casertano}, {Chen}, {Chen}, {Cole}, {Deacon}, {Frenk}, {Fitzsimmons},
  {Gezari}, {Gibbs}, {Goessl}, {Goggia}, {Gourgue}, {Goldman}, {Grant},
  {Grebel}, {Hambly}, {Hasinger}, {Heavens}, {Heckman}, {Henderson}, {Henning},
  {Holman}, {Hopp}, {Ip}, {Isani}, {Jackson}, {Keyes}, {Koekemoer}, {Kotak},
  {Le}, {Liska}, {Long}, {Lucey}, {Liu}, {Martin}, {Masci}, {McLean}, {Mindel},
  {Misra}, {Morganson}, {Murphy}, {Obaika}, {Narayan}, {Nieto-Santisteban},
  {Norberg}, {Peacock}, {Pier}, {Postman}, {Primak}, {Rae}, {Rai}, {Riess},
  {Riffeser}, {Rix}, {R{\"o}ser}, {Russel}, {Rutz}, {Schilbach}, {Schultz},
  {Scolnic}, {Strolger}, {Szalay}, {Seitz}, {Small}, {Smith}, {Soderblom},
  {Taylor}, {Thomson}, {Taylor}, {Thakar}, {Thiel}, {Thilker}, {Unger},
  {Urata}, {Valenti}, {Wagner}, {Walder}, {Walter}, {Watters}, {Werner},
  {Wood-Vasey}, \& {Wyse}}]{Chambers+2016}
{Chambers}, K.~C., {Magnier}, E.~A., {Metcalfe}, N., {et~al.} 2016, arXiv
  e-prints, arXiv:1612.05560.
\newblock \doarXiv{1612.05560}

\bibitem[{{Czerny} {et~al.}(1997){Czerny}, {Bulik}, {Sikora}, \&
  {Vilhu}}]{Czerny+1997}
{Czerny}, B., {Bulik}, T., {Sikora}, M., \& {Vilhu}, O. 1997, arXiv e-prints,
  astro, \dodoi{10.48550/arXiv.astro-ph/9704260}

\bibitem[{{D'Avanzo} {et~al.}(2022){D'Avanzo}, {Ferro}, {Brivio}, {Bernardini},
  {Fugazza}, {Campana}, {Covino}, {D'Elia}, {De Pasquale}, {Malesani},
  {Melandri}, {Palazzi}, {Piranomonte}, {Rossi}, {Sbarufatti}, {Tagliaferri},
  {REM Team}, \& {CIBO Collaboration}}]{Davanzo+2022}
{D'Avanzo}, P., {Ferro}, M., {Brivio}, R., {et~al.} 2022, GRB Coordinates
  Network, 32755, 1

\bibitem[{{DuPont} {et~al.}(2022){DuPont}, {MacFadyen}, \&
  {Zrake}}]{DuPont+2022}
{DuPont}, M., {MacFadyen}, A., \& {Zrake}, J. 2022, \apjl, 931, L16,
  \dodoi{10.3847/2041-8213/ac6ded}

\bibitem[{{Goodman}(1997)}]{Goodman+1997}
{Goodman}, J. 1997, \na, 2, 449, \dodoi{10.1016/S1384-1076(97)00031-6}

\bibitem[{{Granot}(2007)}]{Granot+2007}
{Granot}, J. 2007, in Revista Mexicana de Astronomia y Astrofisica Conference
  Series, Vol.~27, Revista Mexicana de Astronomia y Astrofisica, vol. 27,
  140--165, \dodoi{10.48550/arXiv.astro-ph/0610379}

\bibitem[{{Granot} {et~al.}(1999{\natexlab{a}}){Granot}, {Piran}, \&
  {Sari}}]{Granot+et+al+1999}
{Granot}, J., {Piran}, T., \& {Sari}, R. 1999{\natexlab{a}}, \apj, 527, 236,
  \dodoi{10.1086/308052}

\bibitem[{{Granot} {et~al.}(1999{\natexlab{b}}){Granot}, {Piran}, \&
  {Sari}}]{Granot+Piran+Sari+1999}
---. 1999{\natexlab{b}}, \apj, 527, 236, \dodoi{10.1086/308052}

\bibitem[{{Granot} {et~al.}(1999{\natexlab{c}}){Granot}, {Piran}, \&
  {Sari}}]{Granot+Sari+Piran+1999}
---. 1999{\natexlab{c}}, \apj, 513, 679, \dodoi{10.1086/306884}

\bibitem[{{Gruzinov} \& {Waxman}(1999)}]{Gruzinov+Waxman+1999}
{Gruzinov}, A., \& {Waxman}, E. 1999, \apj, 511, 852, \dodoi{10.1086/306720}

\bibitem[{{Ivezi{\'c}} {et~al.}(2019){Ivezi{\'c}}, {Kahn}, {Tyson}, {Abel},
  {Acosta}, {Allsman}, {Alonso}, {AlSayyad}, {Anderson}, {Andrew}, {Angel},
  {Angeli}, {Ansari}, {Antilogus}, {Araujo}, {Armstrong}, {Arndt}, {Astier},
  {Aubourg}, {Auza}, {Axelrod}, {Bard}, {Barr}, {Barrau}, {Bartlett}, {Bauer},
  {Bauman}, {Baumont}, {Bechtol}, {Bechtol}, {Becker}, {Becla}, {Beldica},
  {Bellavia}, {Bianco}, {Biswas}, {Blanc}, {Blazek}, {Blandford}, {Bloom},
  {Bogart}, {Bond}, {Booth}, {Borgland}, {Borne}, {Bosch}, {Boutigny},
  {Brackett}, {Bradshaw}, {Brandt}, {Brown}, {Bullock}, {Burchat}, {Burke},
  {Cagnoli}, {Calabrese}, {Callahan}, {Callen}, {Carlin}, {Carlson},
  {Chandrasekharan}, {Charles-Emerson}, {Chesley}, {Cheu}, {Chiang}, {Chiang},
  {Chirino}, {Chow}, {Ciardi}, {Claver}, {Cohen-Tanugi}, {Cockrum}, {Coles},
  {Connolly}, {Cook}, {Cooray}, {Covey}, {Cribbs}, {Cui}, {Cutri}, {Daly},
  {Daniel}, {Daruich}, {Daubard}, {Daues}, {Dawson}, {Delgado}, {Dellapenna},
  {de Peyster}, {de Val-Borro}, {Digel}, {Doherty}, {Dubois},
  {Dubois-Felsmann}, {Durech}, {Economou}, {Eifler}, {Eracleous}, {Emmons},
  {Fausti Neto}, {Ferguson}, {Figueroa}, {Fisher-Levine}, {Focke}, {Foss},
  {Frank}, {Freemon}, {Gangler}, {Gawiser}, {Geary}, {Gee}, {Geha}, {Gessner},
  {Gibson}, {Gilmore}, {Glanzman}, {Glick}, {Goldina}, {Goldstein}, {Goodenow},
  {Graham}, {Gressler}, {Gris}, {Guy}, {Guyonnet}, {Haller}, {Harris},
  {Hascall}, {Haupt}, {Hernandez}, {Herrmann}, {Hileman}, {Hoblitt}, {Hodgson},
  {Hogan}, {Howard}, {Huang}, {Huffer}, {Ingraham}, {Innes}, {Jacoby}, {Jain},
  {Jammes}, {Jee}, {Jenness}, {Jernigan}, {Jevremovi{\'c}}, {Johns}, {Johnson},
  {Johnson}, {Jones}, {Juramy-Gilles}, {Juri{\'c}}, {Kalirai}, {Kallivayalil},
  {Kalmbach}, {Kantor}, {Karst}, {Kasliwal}, {Kelly}, {Kessler}, {Kinnison},
  {Kirkby}, {Knox}, {Kotov}, {Krabbendam}, {Krughoff}, {Kub{\'a}nek},
  {Kuczewski}, {Kulkarni}, {Ku}, {Kurita}, {Lage}, {Lambert}, {Lange},
  {Langton}, {Le Guillou}, {Levine}, {Liang}, {Lim}, {Lintott}, {Long},
  {Lopez}, {Lotz}, {Lupton}, {Lust}, {MacArthur}, {Mahabal}, {Mandelbaum},
  {Markiewicz}, {Marsh}, {Marshall}, {Marshall}, {May}, {McKercher}, {McQueen},
  {Meyers}, {Migliore}, {Miller}, {Mills}, {Miraval}, {Moeyens}, {Moolekamp},
  {Monet}, {Moniez}, {Monkewitz}, {Montgomery}, {Morrison}, {Mueller},
  {Muller}, {Mu{\~n}oz Arancibia}, {Neill}, {Newbry}, {Nief}, {Nomerotski},
  {Nordby}, {O'Connor}, {Oliver}, {Olivier}, {Olsen}, {O'Mullane}, {Ortiz},
  {Osier}, {Owen}, {Pain}, {Palecek}, {Parejko}, {Parsons}, {Pease},
  {Peterson}, {Peterson}, {Petravick}, {Libby Petrick}, {Petry},
  {Pierfederici}, {Pietrowicz}, {Pike}, {Pinto}, {Plante}, {Plate}, {Plutchak},
  {Price}, {Prouza}, {Radeka}, {Rajagopal}, {Rasmussen}, {Regnault}, {Reil},
  {Reiss}, {Reuter}, {Ridgway}, {Riot}, {Ritz}, {Robinson}, {Roby}, {Roodman},
  {Rosing}, {Roucelle}, {Rumore}, {Russo}, {Saha}, {Sassolas}, {Schalk},
  {Schellart}, {Schindler}, {Schmidt}, {Schneider}, {Schneider}, {Schoening},
  {Schumacher}, {Schwamb}, {Sebag}, {Selvy}, {Sembroski}, {Seppala}, {Serio},
  {Serrano}, {Shaw}, {Shipsey}, {Sick}, {Silvestri}, {Slater}, {Smith},
  {Smith}, {Sobhani}, {Soldahl}, {Storrie-Lombardi}, {Stover}, {Strauss},
  {Street}, {Stubbs}, {Sullivan}, {Sweeney}, {Swinbank}, {Szalay}, {Takacs},
  {Tether}, {Thaler}, {Thayer}, {Thomas}, {Thornton}, {Thukral}, {Tice},
  {Trilling}, {Turri}, {Van Berg}, {Vanden Berk}, {Vetter}, {Virieux},
  {Vucina}, {Wahl}, {Walkowicz}, {Walsh}, {Walter}, {Wang}, {Wang}, {Warner},
  {Wiecha}, {Willman}, {Winters}, {Wittman}, {Wolff}, {Wood-Vasey}, {Wu},
  {Xin}, {Yoachim}, \& {Zhan}}]{Ivezi+2019}
{Ivezi{\'c}}, {\v{Z}}., {Kahn}, S.~M., {Tyson}, J.~A., {et~al.} 2019, \apj,
  873, 111, \dodoi{10.3847/1538-4357/ab042c}

\bibitem[{{Katz}(1994)}]{Katz+1994}
{Katz}, J.~I. 1994, \apjl, 432, L107, \dodoi{10.1086/187523}

\bibitem[{{Kochanek} {et~al.}(2017){Kochanek}, {Shappee}, {Stanek}, {Holoien},
  {Thompson}, {Prieto}, {Dong}, {Shields}, {Will}, {Britt}, {Perzanowski}, \&
  {Pojma{\'n}ski}}]{Kochanek+2017}
{Kochanek}, C.~S., {Shappee}, B.~J., {Stanek}, K.~Z., {et~al.} 2017, \pasp,
  129, 104502, \dodoi{10.1088/1538-3873/aa80d9}

\bibitem[{{Kumar} \& {Zhang}(2015)}]{Kumar+zhang+2015}
{Kumar}, P., \& {Zhang}, B. 2015, \physrep, 561, 1,
  \dodoi{10.1016/j.physrep.2014.09.008}

\bibitem[{{Mandarakas} {et~al.}(2023){Mandarakas}, {Blinov}, {Aguilera-Dena},
  {Romanopoulos}, {Pavlidou}, {Tassis}, {Antoniadis}, {Kiehlmann}, {Lychoudis},
  \& {Tsemperof Kataivatis}}]{Mandarakas+2023}
{Mandarakas}, N., {Blinov}, D., {Aguilera-Dena}, D.~R., {et~al.} 2023, \aap,
  670, A144, \dodoi{10.1051/0004-6361/202244802}

\bibitem[{{Mooley} {et~al.}(2018){Mooley}, {Deller}, {Gottlieb}, {Nakar},
  {Hallinan}, {Bourke}, {Frail}, {Horesh}, {Corsi}, \&
  {Hotokezaka}}]{Mooley+2018}
{Mooley}, K.~P., {Deller}, A.~T., {Gottlieb}, O., {et~al.} 2018, \nat, 561,
  355, \dodoi{10.1038/s41586-018-0486-3}

\bibitem[{{Nakar} {et~al.}(2009){Nakar}, {Ando}, \& {Sari}}]{Nakar+2009}
{Nakar}, E., {Ando}, S., \& {Sari}, R. 2009, \apj, 703, 675,
  \dodoi{10.1088/0004-637X/703/1/675}

\bibitem[{{Piran}(2004)}]{Piran+2004}
{Piran}, T. 2004, Reviews of Modern Physics, 76, 1143,
  \dodoi{10.1103/RevModPhys.76.1143}

\bibitem[{{Rhoads}(1999)}]{Rhoads+1999}
{Rhoads}, J.~E. 1999, \apj, 525, 737, \dodoi{10.1086/307907}

\bibitem[{{Rossi} {et~al.}(2002){Rossi}, {Lazzati}, \& {Rees}}]{Rossi+2002}
{Rossi}, E., {Lazzati}, D., \& {Rees}, M.~J. 2002, \mnras, 332, 945,
  \dodoi{10.1046/j.1365-8711.2002.05363.x}

\bibitem[{{Rybicki} \& {Lightman}(1979)}]{Rybciki+Lightman+1979}
{Rybicki}, G.~B., \& {Lightman}, A.~P. 1979, {Radiative processes in
  astrophysics}

\bibitem[{{Sari}(1999)}]{Sari+1999}
{Sari}, R. 1999, \apjl, 524, L43, \dodoi{10.1086/312294}

\bibitem[{{Sari} \& {Esin}(2001)}]{Sari+Esin+2001}
{Sari}, R., \& {Esin}, A.~A. 2001, \apj, 548, 787, \dodoi{10.1086/319003}

\bibitem[{{Sari} {et~al.}(1999){Sari}, {Piran}, \&
  {Halpern}}]{Sari+Piran+Halpern+1999}
{Sari}, R., {Piran}, T., \& {Halpern}, J.~P. 1999, \apjl, 519, L17,
  \dodoi{10.1086/312109}

\bibitem[{{Sari} {et~al.}(1998){Sari}, {Piran}, \&
  {Narayan}}]{Sari+Piran+Narayan+1998}
{Sari}, R., {Piran}, T., \& {Narayan}, R. 1998, \apjl, 497, L17,
  \dodoi{10.1086/311269}

\bibitem[{{Shappee} {et~al.}(2014){Shappee}, {Prieto}, {Grupe}, {Kochanek},
  {Stanek}, {De Rosa}, {Mathur}, {Zu}, {Peterson}, {Pogge}, {Komossa}, {Im},
  {Jencson}, {Holoien}, {Basu}, {Beacom}, {Szczygie{\l}}, {Brimacombe},
  {Adams}, {Campillay}, {Choi}, {Contreras}, {Dietrich}, {Dubberley},
  {Elphick}, {Foale}, {Giustini}, {Gonzalez}, {Hawkins}, {Howell}, {Hsiao},
  {Koss}, {Leighly}, {Morrell}, {Mudd}, {Mullins}, {Nugent}, {Parrent},
  {Phillips}, {Pojmanski}, {Rosing}, {Ross}, {Sand}, {Terndrup}, {Valenti},
  {Walker}, \& {Yoon}}]{Shappee+2014}
{Shappee}, B.~J., {Prieto}, J.~L., {Grupe}, D., {et~al.} 2014, \apj, 788, 48,
  \dodoi{10.1088/0004-637X/788/1/48}

\bibitem[{{Stanek} {et~al.}(2003){Stanek}, {Matheson}, {Garnavich}, {Martini},
  {Berlind}, {Caldwell}, {Challis}, {Brown}, {Schild}, {Krisciunas}, {Calkins},
  {Lee}, {Hathi}, {Jansen}, {Windhorst}, {Echevarria}, {Eisenstein}, {Pindor},
  {Olszewski}, {Harding}, {Holland}, \& {Bersier}}]{Stanek+03}
{Stanek}, K.~Z., {Matheson}, T., {Garnavich}, P.~M., {et~al.} 2003, \apjl, 591,
  L17, \dodoi{10.1086/376976}

\bibitem[{{Taylor} {et~al.}(2004){Taylor}, {Frail}, {Berger}, \&
  {Kulkarni}}]{Taylor+2004}
{Taylor}, G.~B., {Frail}, D.~A., {Berger}, E., \& {Kulkarni}, S.~R. 2004,
  \apjl, 609, L1, \dodoi{10.1086/422554}

\bibitem[{{van Eerten} {et~al.}(2010){van Eerten}, {Leventis}, {Meliani},
  {Wijers}, \& {Keppens}}]{vanEerten+2010}
{van Eerten}, H.~J., {Leventis}, K., {Meliani}, Z., {Wijers}, R.~A.~M.~J., \&
  {Keppens}, R. 2010, \mnras, 403, 300,
  \dodoi{10.1111/j.1365-2966.2009.16109.x}

\bibitem[{{Williams} {et~al.}(2023){Williams}, {Kennea}, {Dichiara},
  {Kobayashi}, {Iwakiri}, {Beardmore}, {Evans}, {Heinz}, {Lien}, {Oates},
  {Negoro}, {Cenko}, {Buisson}, {Hartmann}, {Jaisawal}, {Kuin}, {Lesage},
  {Page}, {Parsotan}, {Pasham}, {Sbarufatti}, {Siegel}, {Sugita}, {Younes},
  {Ambrosi}, {Arzoumanian}, {Bernardini}, {Campana}, {Capalbi}, {Caputo},
  {D'Ai}, {D'Avanzo}, {D'Elia}, {De Pasquale}, {Eyles-Ferris}, {Ferrara},
  {Gendreau}, {Gropp}, {Kawai}, {Klingler}, {Laha}, {Melandri}, {Mihara},
  {Moss}, {O'Brien}, {Osborne}, {Palmer}, {Perri}, {Serino}, {Sonbas},
  {Stamatikos}, {Starling}, {Tagliaferri}, {Tohuvavohu}, {Zane}, \&
  {Ziaeepour}}]{Williams+23}
{Williams}, M.~A., {Kennea}, J.~A., {Dichiara}, S., {et~al.} 2023, arXiv
  e-prints, arXiv:2302.03642, \dodoi{10.48550/arXiv.2302.03642}

\bibitem[{{Yonetoku} {et~al.}(2011){Yonetoku}, {Murakami}, {Gunji}, {Mihara},
  {Toma}, {Sakashita}, {Morihara}, {Takahashi}, {Toukairin}, {Fujimoto},
  {Kodama}, {Kubo}, \& {IKAROS Demonstration Team}}]{Yonetoku+2011}
{Yonetoku}, D., {Murakami}, T., {Gunji}, S., {et~al.} 2011, \apjl, 743, L30,
  \dodoi{10.1088/2041-8205/743/2/L30}

\bibitem[{{Zhang} \& {M{\'e}sz{\'a}ros}(2002)}]{Zhang+2002}
{Zhang}, B., \& {M{\'e}sz{\'a}ros}, P. 2002, \apj, 571, 876,
  \dodoi{10.1086/339981}

\end{thebibliography}
\bibliographystyle{aasjournal}



\end{document}